# Mass Parameterizations and Predictions of Isotopic Observables

S.R. Souza[1], P. Danielewicz[2], S. Das Gupta[3], R. Donangelo[1], W. A. Friedman[4], W.G. Lynch[2], W.P. Tan[2*] , M.B. Tsang[2#]

[1]*Instituto de Física, Universidade Federal do Rio de Janeiro, Cidade Universitária, CP 68528, 21941-972 Rio de Janeiro, Brazil*

[2]*National Superconducting Cyclotron Laboratory and Department of Physics and Astronomy, Michigan State University, East Lansing, MI 48824, USA*

[3]*Physics Department., McGill University, 3600 University Street, Montreal, Canada H3A 2T8*

[4]*Department of Physics, University of Wisconsin, Madison, WI 53706, USA*

\* *present address: University of Notre Dame, South Bend, IN*

# *e-mail address: tsang@nscl.msu.edu*

**Abstract**

We discuss the accuracy of mass models for extrapolating to very asymmetric nuclei and the impact of such extrapolations on the predictions of isotopic observables in multifragmentation. We obtain improved mass predictions by incorporating measured masses and extrapolating to unmeasured masses with a mass formula that includes surface symmetry and Coulomb terms. We find that using accurate masses has a significant impact on the predicted isotopic observables.

Masses are critically important parameters in statistical models. They define the baryon number, the minimum energies of each decay mode, and enter exponentially into the Boltzmann factors that dictate the relevant yields [1-3] or emission rates [4]. Most statistical models utilize measured masses for frequently emitted species such as neutron, hydrogen and helium isotopes [1-7]; however, mass formulae must be employed to predict unknown masses. These unknown masses typically have unusual magnitude of isospin asymmetries, and their masses can influence the yields of their stable counterparts more significantly than is often realized. Here, we explore the interplay between these mass assumptions and predicted isotopic distributions within the context of an equilibrium model for multifragmentation [2, 5-11]. Over the past two decades, there have been many different variations of the statistical multifragmentation models first described in details in ref. [6]. To avoid confusions, we label relevant versions of the SMM codes with the associated references to be discussed here; the two SMM codes originated from Copenhagan are SMM85 [6] and SMM95 [2, 5]; ISMM [10,11] is the improved SMM code with empirical masses and level densities based on the microcanonical SMM85; ISMM_McGill which is used in the present work, is the canonical SMM code using the recursive relations developed by Das Gupta [8,9] and incorporates the improvements developed in ISMM. The isotope distributions produced by ISMM_McGill are similar to those predicted by ISMM described in details in Ref. [10] and [11].

We begin by discussing some of the deficiencies of mass formulae that are used in statistical models and some remedies relevant to the description of very asymmetric nuclei. Many mass formulae owe their form to the semi-empirical or Liquid Drop Mass (LDM) parametrization introduced by Weizsacker [12-15]. Such formulae approximate the nuclear mass $M(A,Z)$ by

$$M(A,Z) = N \cdot m_n + Z \cdot m_p - BE_{LDM}(A,Z)/c^2$$

where

$$BE_{LDM}(A,Z) = a_v A - a_s A^{2/3} - a_c Z^2/A^{1/3} + a_p A^{-1/2} - a_{sym}(N-Z)^2/A \qquad (1)$$

and $m_n$ ($m_p$) are the neutron (proton) masses; N, Z and A are the neutron, proton and nucleon numbers; and $a_v$, $a_s$, $a_c$, and $a_{sym}$ are the coefficients of volume, surface, Coulomb, and symmetry in the liquid drop model. The value of the pairing term $a_p$ is 0 for odd A, positive for even N and Z and negative for odd N and Z. The A-dependence of $a_v$, $a_s$, $a_c$ and $a_{sym}$ follows from the A dependence of the geometry of a well bound spherical nucleus, but the A dependence of the pairing term does not. Other forms of the pairing terms can be found in the literature [13-16]. Many different values for the coefficients used in Eq. (1) have been reported; some typical values used in the SMM models of ref. [2] and [8] are given in Table I. Other SMM models used in the literature reported different parameter sets [17].

Statistical models typically utilize mass formulae such as Eq. (1) for many, if not all, of the heavier masses. Surprisingly, the mass formulae that are utilized are often not particularly accurate. For example, the upper panel (a) of Fig. 1 shows the differences $\Delta BE = BE_{LDM}(A,Z) - BE_{EXP}(A,Z)$ between the *total* binding energies $BE_{LDM}(A,Z)$ employed by SMM95 [2] and the empirical binding energies, $BE_{EXP}(A,Z)$, tabulated by Audi and Wapstra [18] For the heavier masses, there are discrepancies, which attain values as large as 40 MeV. Even for light charged particles, the calculated masses can be off by over 20 MeV. In terms of the binding energy per nucleon, these discrepancies for the heaviest masses are less than 150 keV and may appear small. However, it is the total binding energy and not the binding energy per nucleon, that enters into statistical models [2, 5, 8]. For temperatures of the order of 5 MeV, typically assumed in these models [8, 19, 20], such discrepancies correspond to changes in the Boltzmann factor and in the production probabilities for these nuclei that are of the order of $\exp(\Delta BE/T) \approx 400$ - too large to be ignored

Advances in experimental measurements have provided high quality isotopically resolved data for neutron-rich systems [21-23]. Multifragmentation calculations for such systems require values for unmeasured masses of nuclei with neutron numbers N and charge numbers Z that lie very far from the valley of stability. Mass formulae of the form in Eq. (1) assume the symmetry coefficient $a_{sym}$ to be independent of nucleon number. However, it appears that there should be a nucleon number dependence of the symmetry coefficient, reflecting the density dependence of the asymmetry term of the nuclear equation of state [1, 24,25]. To incorporate this, both the asymmetry and the Coulomb terms in the mass formula should be separated into bulk and surface contributions [1, 14, 24, 25]. These surface symmetry and Coulomb terms are required for very neutron-rich nuclear matter because the surfaces can accumulate a significant fraction of asymmetry [25,26].

To incorporate these surface energy terms, we adopt for simplicity the parametrization of the improved LDM (ILDM) formula of ref. [14, 24]:

$$BE_{ILDM}(A,Z) = a_v[1-k((A-2Z)/A)^2]A - a_s[1-k((A-2Z)/A)^2]A^{2/3} - a_c Z^2/A^{1/3} + a_p A^{-1/2} + c_d Z^2/A, \quad (2)$$

Here, the extra Coulomb term $c_d Z^2/A$, neglected in most models, takes into account corrections to the Coulomb energy associated with the diffuseness of the nuclear surface. The symmetry terms in Eq. (2) can be regrouped in a form similar to Eq.(1). From this one can identify an effective total asymmetry coefficient $a_{sym}'$ of Eq. (2) that includes the contribution from the surface and is dependent on A

$$a_{sym}' = k(a_v - a_s A^{-1/3}) \quad (3)$$

The parameters of Eq. (2) listed in the fourth row in Table I correspond to the best fit of the experimental data for A≥5 in the Audi-Wapstra table [17]. The fit includes 2920 experimental masses. Figure 1b shows the difference between the binding energies calculated with the best fit parameters of ILMD and the Audi-Wapstra table [17] . The disagreement is much reduced relative to the comparison in Fig. 1a; the remaining deviations arise mainly from shell effect corrections.

To achieve the most accurate treatment of the masses, we employ the tabulated masses in Audi-Wapstra table [17] when they are known. However we still need to compute the unknown masses for some nuclei, most of which have extreme proton to neutron composition. We adopt for simplicity, a procedure in which we compute the average shift of the ILDM formula from the empirical values near the extremes of the BE(A,Z) vs Z relationship at fixed neutron number. This shift, $\Delta_N$, is then subtracted from the prediction of the ILDM formula:

$$BE_{extrap}(A,Z) = BE_{ILDM}(A,Z) - \Delta_n, \quad (4)$$

where

$$\Delta_n = (1/n)\Sigma_i (BE_{ILDM}(A_i, Z_i) - BE_{recomm}(A_i, Z_i)),  \qquad (5)$$

and n=3 is the number of points taken before the right or left end of the curve. For example, $^{46}$Fe is the heaviest and $^{29}$F is the lightest isotone for N=20 listed in the Audi-Wapstra table [17]. To predict the binding energy for N=20 isotones heavier than $^{46}$Fe, we use the masses of $^{44}$Cr, $^{45}$Mn and $^{46}$Fe and Eqs. 4 and 5. Similarly, we compute $\Delta_n$ from the masses of $^{29}$F, $^{30}$Ne and $^{31}$Na to predict the masses of N=20 isotones lighter than $^{29}$F.

To check this extrapolation procedure, we performed a similar analysis in which we treated the masses of the lightest and heaviest nuclei with fixed neutron number N in the Audi-Wapstra table [17] as unknown. We then predict the masses of these isotones using Eqs. 4 and 5. Using the previous example, this means we use shifts $\Delta_n$ obtained from $^{43}$V, $^{44}$Cr, and $^{45}$Mn to predict $^{46}$Fe, and $\Delta_n$ obtained from $^{30}$Ne, $^{31}$Na and $^{32}$Mg to predict $^{29}$F. In Figure 2 we show the differences between the calculated and the empirical masses for the extreme ends of the isotone distributions as solid points [27]. We contrast this with the open squares, which denote the corresponding differences between the empirical masses and those calculated from the ILDM without this correction. Since this extrapolation is applied only to unknown masses, from this comparison, we estimate that our final procedure including the corrections of Eqs. 4 and 5 provides masses with accuracies about 1-2 MeV for nuclei just outside the Audi-Wapstra table. These extrapolations become less accurate with decreasing Z.

Now, we examine the sensitivity of the isotopic distributions predicted by the improved multifragmentation model, ISMM_McGill, to the masses used. Major improvement includes the incorporation of empirical binding energies and level densities [10, 11] in the multifragmentation stage. We compare predictions obtained by using the "standard" LDM mass formula (Eq. 1 with parameters listed in the first row of Table 1) to predictions using empirical masses supplemented by the ILDM mass formula described above.

Following Ref. [28], we perform calculations for two systems with source charge of $Z_0$ =75 and masses of $A_0$=168 ($N_0/Z_0$=1.24) and $A_0$=186 ($N_0/Z_0$=1.48). These two sources correspond to estimates of the prefragments remaining after pre-equilibrium emission in central $^{112}$Sn+$^{112}$Sn and $^{124}$Sn+$^{124}$Sn collisions, respectively, at an incident energy of E/A=50 MeV. We assume a break up density of $\rho_0/6$ where $\rho_0$=0.16 fm$^{-3}$ and a temperature of 4.7

MeV. This latter value corresponds to the average "temperature" of fragments produced in the corresponding microcanonical ISMM models, at a total prefragment excitation energy of 5 MeV per nucleon [10].

The open data points in the left panels of Fig. 3 show the primary oxygen isotope distributions (before secondary decay) when the standard LDM masses are used for sources of $A_0$=186 (upper panel) and $A_0$=168 (lower panel). The predicted distributions are approximately Gaussian. Using the combination of empirical and extrapolated ILDM masses yields the primary distributions given by solid points in the left panels of Fig. 3. These latter distributions are much wider and display a notably higher production of the neutron-rich isotopes in the tails of the isotope distribution. This feature occurs for both the larger and more neutron-rich source (upper panel) and the smaller and more neutron-deficient source (lower panel). Similar widening of the isotopic distributions also occurs for the other elements. Thus the standard LDM masses used in most SMM calculations provide primary distributions that are much narrower and more neutron-deficient than those calculated when more realistic masses are used.

Since the experimental isotopic distributions reflect the particle decay of excited particle unstable fragments, one should examine the isotope distributions after the sequential decays. There are many models that simulate the effects of sequential decays. In this work, we choose two sequential decays for comparison. The most sophisticated sequential decay algorithm included in ISMM [10] uses the empirical and the ILDM masses with empirical level densities. This decay code has been developed at the Michigan State University over the years [29,10] and is called the MSU_DECAY [10]. The solid points in the right panels of Fig. 3 denote the final oxygen isotopic distributions obtained from ISMM_McGill with MSU_DECAY calculation. In this calculation (Calc I), both the multifragmentation calculations and the secondary decays calculations use the empirical and the ILDM masses and empirical level densities self-consistently. The dashed lines in the right panels (calculation II) indicate the final oxygen isotopic distributions obtained from ISMM_McGill calculation with LDM masses (open points in left panels) with MSU_DECAY. Thus calculation II is not self-consistent; different masses are used to calculate the primary (LDM masses) and secondary decay calculations (empirical and ILDM masses). Even though the primary distributions of Calc I and Calc II are very different, the final distributions after the sequential decays are quite similar. Nevertheless, differences of the order of a factor of two are observed between calc. I (solid points) and calc. II (dashed lines).

The secondary decay prescriptions used in most SMM [2, 7] codes adopt the evaporation and fermi break-up [30] as the decay process. We call this decay code the SMM_DECAY. The masses used in such calculations include both empirical and calculated masses. Empirical masses are used for most light nuclei with A<20 and parametrization of the masses for all the others. The open points (calculation III) correspond to the final distributions when this sequential decay algorithm is applied to the primary calculations obtained using LDM masses for A>4. The difference between Calc. I (solid points) and Calc. III (open points) is large. The predictions from Calc. III should be similar to results obtained from prior SMM calculations such as the version used in ref. [20]. Thus, the different mass assumptions as well as the sequential decays could be an important factor in explaining why fragment distributions that use codes similar to SMM95 are much narrower and under-predict the production of neutron-rich nuclei.

Recent studies suggest that detailed comparison between reactions at the same temperature or excitation energy, but at different proton to neutron composition, can be made using the isoscaling relationship [22,31,32]

$$Y_2(N,Z)/Y_1(N,Z) = C\exp(\alpha N+\beta Z), \qquad (6)$$

where $C$, $\alpha$, and $\beta$, are fitting parameters. The subscripts 1 and 2 refer to the two sources with different isospin composition, with source 2 normally referring to the more neutron-rich source. In the present work, 1 and 2 denote sources with nucleon and charge numbers, ($A_o$, $Z_o$) corresponding to (168, 75) and (186, 75), respectively. Primary fragments produced in grand canonical, canonical and microcanonical statistic multifragmentation models generally obey isoscaling [28]. The extracted isoscaling parameters depend strongly on the isospin asymmetry of the source, but they may also depend on the isospin dependence of the masses used. Indeed, different isoscaling fitting parameter sets, $C$, $\alpha$, and $\beta$, are extracted depending upon whether the masses are obtained from the LDM or from the empirical values plus the shell corrected ILDM calculations. These isoscaling parameters are listed in columns 3-5 of Table II. The absolute values of the relevant $\alpha$ and $\beta$ parameters, resulting from fitting the calculations that use the LDM masses are higher than those that use the empirical plus ILDM masses.

Depending on which masses and the decay mechanisms are employed in the sequential feeding algorithm, different isotope distributions will result. Figure 4 shows the isotope ratios and isoscaling fits for Calc I, II and III. The open and closed points are the predicted isotope ratios as a function of N for odd (Z=3, 5, 7) and even (Z=4, 6, 8) charge elements respectively. The solid and dashed lines correspond to the best fit lines. The isoscaling fits (lines) from calc. III shown in the rightmost panel of Figure 3 vary nonstatistically with respect to the predictions (symbols); indicating that isoscaling is not well obeyed by the schematic secondary decay approach of ref. [28]. In comparison, better fits are obtained for Calc I (left panel) and Calc II (middle panel), for which the sequential decays are calculated using the empirical ILDM masses and empirical level densities. Due to the similarity in the final distributions of Calc I and II, the $\alpha$ values obtained are similar even though the $\alpha$ values from the primary fragment distributions are quite different. However, there are significant differences in the final value for $\alpha$ and $\beta$ between the fully empirically based Calc. I and Calc. III, indicating that precise treatments of the mass values and sequential decays should be implemented within equilibrium statistical multifragmentation models before they can be used with confidence to describe isoscaling observables.

It is interesting to note that corrections to $\alpha$ from secondary decay in Calc. I where consistent values for the masses are used in the primary and secondary decay calculations are smaller than for Calc. II where mass values for primary and secondary decay stages are different. The situation is less clear for $\beta$, which is affected by Coulomb interactions in the freezeout configuration that influence the primary yields but do not enter into the calculation of the secondary decays. Large differences have been observed between isoscaling parameters extracted from primary fragments produced by the dynamical Stochastic Mean Field (SMF) model and the corresponding parameters after decay [23]. It would be interesting to know whether such differences may be caused in part by discrepancies between the SMF masses and the ones used in secondary decay as was observed in the case of Calc. II above.

In summary, recent experimental advances in measuring isotope distributions and the improvement of multifragmentation models suggest that accurate fragment masses should be incorporated into these models to provide accurate comparison between data and theoretical predictions. The effect of inaccuracies in the mass parametrization upon isotopic observables should not be limited to the SMM approach, but should apply to all statistical and dynamical models of fragment production.

This work is supported by the National Science Foundation under Grant Nos. PHY-01-10253, PHY-00-70818, PHY- 00-70161, INT-9908727 and by the contract No. 41.96.0886.00 of MCT/FINEP/CNPq (PRONEX).

**Tables:**

Table I : List of parameters used in the simple LDM (Eq. 1) and the ILDM (Eq. 2).

| Parameter /Model | $a_v$ | $a_s$ | $a_c$ | $a_{sym}$ | $C_d$ | $a_v k$ | $a_s k$ | $a_p$ |
|---|---|---|---|---|---|---|---|---|
| LDM [2] | 16.0 | 18.0 | 0.72 | 23.0 | n/a | n/a | n/a | 0 |
| LDM [8] | 15.8 | 18.0 | 0.72 | 23.5 | n/a | n/a | n/a | |
| ILDM | 15.6658 | 18.9952 | 0.72053 | n/a | 1.74859 | 27.7976 | 33.7053 | 10.857 |

Table II : Best fit isoscaling parameters. The calculations of the primary and secondary distributions are labeled by the mass formulae that are used.

| Calc. | Primary | C (before) | α (before) | β(before) | Decay | C(after) | α (after) | β((after) |
|---|---|---|---|---|---|---|---|---|

| I | ILDM | 1.1349 | 0.4847 | -0.6511 | ILDM | 0.8501 | 0.459 | -0.481 |
|---|------|--------|--------|---------|------|--------|-------|--------|
| II | LDM | 1.1477 | 0.6233 | -0.8478 | ILDM | 0.9175 | 0.433 | -0.501 |
| III | LDM | 1.1477 | 0.6233 | -0.8478 | LDM-like | 0.4754 | 0.592 | -0.572 |

**Figures Captions:**

**Figure 1:** Deviation of calculated binding energies from empirical binding energies [17]. The calculated masses are obtained using (a) Eq 1 with parameters of ref [2] (see Table 1) and (b) Eq. 2 with the best fit values listed in Table I labeled ILMD.

**Figure 2:** Deviation of calculated binding energies from empirical values [17] at the extremes of the BE(N,Z) vs. Z curve. The open squares correspond to the calculated mass using Eq. 2 whereas the full circles represent results obtained with the extrapolated procedures of Eq. 4 and 5 as discussed in the text.

**Figure 3:** Oxygen isotope yields from the improved ISMM_McGill code using LDM parameters of Ref. [2] (open circles) and empirical masses supplemented by ILDM masses (closed circles). The top and bottom panels correspond to different sources (A=186, Z=75) and (A=168, Z=75) respectively. Primary yields are plotted on the left panels and the yields after sequential decays are plotted in the right panels. See text for details.

**Figure 4:** Isoscaling for isotopes with Z=3-8 obtained from the Calc. I, II and III listed in Table II and described in the text. The open and closed circles are predicted ratios and the dashed and solid lines are best fits using Eq. 6.